# Microstructural evolution throughout the structural transition in 1111 oxy-pnictides

A. Martinelli[1], A. Palenzona[1,2], M.Putti[1,3], C.Ferdeghini[1]

[1] SPIN-CNR, corso Perrone 24, 16152 Genova - Italy

[2] Dept. of Chemistry and Industrial Chemistry, University of Genoa, via Dodecaneso 31, 16146 Genova - Italy

[3] Dept. of Physics, University of Genoa, via Dodecaneso 33, 16146 Genova - Italy

**Abstract**

The microstructural evolution throughout the first order tetragonal to orthorhombic structural transition is analyzed by powder diffraction analysis for two different systems belonging to the class of compounds referred to as 1111 oxy-pnictides: $(La_{1-y}Y_y)FeAsO$ and $SmFeAs(O_{1-x}F_x)$. Both systems are characterized by a similar behaviour: on cooling microstrain along the tetragonal $hh0$ direction takes place and increases as the temperature is decreased. Just above the structural transition microstrain reaches its maximum value and then is abruptly suppressed by symmetry breaking. No volume discontinuity throughout the first order transition is observed and a group-subgroup relationship holds between the tetragonal and the orthorhombic structures, thus suggesting that orbital ordering drives symmetry breaking. Microstrain reflects a distribution of lattice parameters in the tetragonal phase and explains the occurrence of anisotropic properties commonly attributed to nematic correlations; in this scenario the nematic behaviour is induced by the tendency towards ordering of Fe orbitals.

## 1. Introduction

1111-type oxy-pnictides belong to the class of materials referred to as Fe-based superconductors and are characterized by the general formula $RE$FeAsO ($RE$: rare earth); they attracted many investigations after the discovery that F-substitution at O site (electron doping) can induce a relatively high superconducting transition temperature ($T_c$) in LaFeAsO.[1]. At room temperature these compounds crystallize in the tetragonal system (*P4/nmm* space group), whereas on cooling undergo a structural transition at $T_{T-O}$ producing an orthorhombic *Cmme* structure. It is commonly reported that symmetry breaking is suppressed when the F content exceeds a critical value,[2,3,4,5,6] but recently we ascertained that the structural effect of F-substitution rests in a reduction of the orthorhombic distortion and symmetry breaking is still active even at optimal doping,[7] similarly to what is observed in the homologous hole-doped samples.[8]

The origin of the structural transition and its interplay with magnetism is still debated; in fact undoped 1111-type compounds exhibit magnetic ordering at $T_M$, a few tens of degree lower than $T_{T-O}$. Earlier investigations claimed a coupled suppression of both symmetry breaking and magnetic ordering for critical values of F-substitution, with a consequent arising of superconductivity. Conversely more careful analyses revealed that in several cases a superconductive ground state can take place even in doped orthorhombic compounds. On the other hand magnetic order has been observed only after the establishment of the orthorhombic symmetry; hence the orthorhombic structure must take place for the occurrence of static magnetism, but not *viceversa*. Among the different mechanisms proposed to explain the occurrence of symmetry breaking magnetic frustration, nematic correlations and orbital ordering are the most credited.

Magnetic frustration can be originated by competing antiferromagnetic near-neighbours and next-near-neighbours interactions between local Fe moments and is relieved by the orthorhombic distortion; in this scenario the structural transition is thus magnetically driven.[9,10] At present experimental data indicate that magnetic ordering can only take place within the orthorhombic structure, not in the tetragonal one, and that $T_{T-O} > T_M$, whatever the composition; in other words spin ordering can not occur in the tetragonal phase. Conversely the structural transition can take place even when magnetic ordering is completely missing.[7,8] In this context it is quite surprising that magnetic interactions could drive the structural transition, but not develop a magnetic structure after the establishment of the orthorhombic structure.

An electronic nematic phase is marked by the breaking of the electronic symmetry compared to that of the underlying lattice[11] and therefore a nematic phase can be detected by the presence of anisotropic properties that locally breaks the lattice symmetry. Magnetic correlations has been argued to play a primary role even for a nematic phase and, in this scenario, the structural transition can be driven by the coupled electronic and magnetic nematic interactions.[12] However the origin of the nematic state has not yet clarified.

Orbital ordering mechanism involves the ordering of $3d$ Fe orbitals and several theoretical calculations suggest that is responsible for both lattice distortion and magnetic ordering.[13,14,15,16] Recently the concept of nematic orbital order has been introduced; in particular theoretical investigations revealed that the anisotropy observed above the structural transition can be originated by orbital ordering, irrespective of whether long-range magnetic order is present or not.[17] Further the occurrence of a nematic orbital order has been proposed; in this scenario magnetic order always follows or is coincident with symmetry breaking.[18]

In order to elucidate the nature of the structural transition we carried out a careful structural and microstructural analysis of the SmFeAs($O_{1-x}F_x$) system; some results of this investigation have been previously published[7] and are here discussed in more detail. In this paper we will focus our attention on the microstructural properties of these compounds and on their evolution throughout the structural transition. The results of this analysis are compared, supported and integrated with those obtained for the homologous ($La_{1-y}Y_y$)FeAsO system, previously investigated.[19] In this system Y substitution induces chemical pressure; as a result symmetry breaking and magnetic ordering temperatures progressively reduce by similar amounts with increasing Y content.

## 2. Experimental

Details concerning the preparation, the resistive and magnetic characterization of both SmFeAs(O$_{1-x}$F$_x$) and (La$_{1-y}$Y$_y$)FeAsO sample series were previously described.[7,19,20,21,22] The samples underwent two heating treatments at high temperature, the former to synthesize the desired phase, the latter, after intermediate grinding, to homogenize the chemical composition. (La$_{1-y}$Y$_y$)FeAsO samples were synthesized in evacuated quartz flasks, whereas SmFeAs(O$_{1-x}$F$_x$) samples in sealed Ta crucibles to reduce F losses, originated by the partial reaction of this element with quartz flask.

Figure 1 shows the resistivity curves (normalized at $r$@300 K) of the SmFeAs(O$_{1-x}$F$_x$) and (La$_{1-y}$Y$_y$)FeAsO sample series analyzed in this paper. (La$_{1-y}$Y$_y$)FeAsO compounds (right panel) exhibit the usual behaviour of the resistivity in the 1111 parent compounds with a maximum followed by a sharp drop with an inflection point corresponding to the maximum of d$r$/d$T$, related to the occurrence of the magnetic transition.[19] An inspection of the resistive curves of the SmFeAs(O$_{1-x}$F$_x$) series with increasing electron doping (left panel) shows that the $x$ = 0.00 and $x$ = 0.05 samples exhibit the maximum, signature of the magnetic transition, the $x$ = 0.075 and 0.085 samples exhibit both a bump, residual signature of the magnetic transition, and the superconducting transition; only the superconducting transition occurs for samples with $x \geq 0.1$.

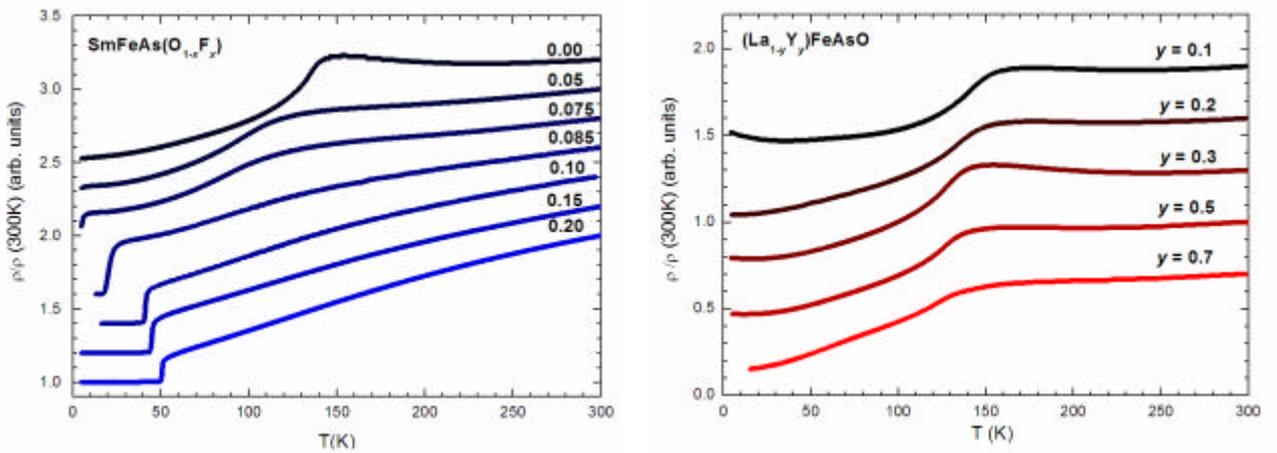

Figure 1: Resistivity curves (normalized at $r$@300 K) of the SmFeAs(O$_{1-x}$F$_x$) and (La$_{1-y}$Y$_y$)FeAsO samples; curves are arbitrarily shifted along the $y$ axis for clarity.

The magnetic behaviour of the samples belonging to the SmFeAs($O_{1-x}F_x$) series was also investigated by means of muon spin rotation (μ-SR) analysis:[23] in particular the occurrence of a static magnetic ordering at the Fe plane was ascertained in the specimens with $x$ up to 0.085, whereas for SmFeAs($O_{0.90}F_{0.10}$) magnetic ordering is completely suppressed, in agreement with resistivity measurements. Hence specimens with $x$ = 0.00, 0.05, 0.075, 0.10 and 0.20, well representing the transition from the magnetic ground state to a fully superconducting state, were analyzed by high resolution synchrotron radiation powder diffraction (SRPD) analysis, carried out at the BM1B beamline of the European Synchrotron Radiation Facility (ESRF) (Grenoble - France). Data were collected between 90 K and 290 K ($\lambda$ = 0.50663 Å; step = 0.004° 2θ). Unfortunately, the experimental setup for SRPD data acquisition prevents to register the temperature at which each analytical file is acquired; hence the temperature was estimated on a chronological basis, thus suffering of a few uncertainties that can be estimated as 10 K at maximum.

Two kinds of SRPD data were collected: in the former case full profile diffraction patterns (angular range = 1.0 – 55° 2θ) were collected at selected temperatures in order to carry out structural refinements applying the Rietveld method using the program FullProf.[24] For these refinements the peak shape was modelled by a Thompson-Cox-Hastings pseudo-Voigt convoluted with axial divergence asymmetry function coupled with an instrumental file in order to evaluate structural microstrain. In the final cycle the following parameters were refined: the scale factor; the zero point of detector; the background (3 parameters of the 5$^{th}$ order polynomial function); the unit cell parameters; the atomic site coordinates in general positions; the isotropic displacement parameter; the anisotropic strain parameters, by means of which Williamson-Hall plots were obtained to investigate the microstructure of the samples.

In the latter case high statistic – high resolution thermo-diffractograms of the 110 diffraction peak (angular range = 10.32 – 10.60° 2θ) were acquired on cooling (30 K/h) in a continuous scanning mode to evaluate the temperature at which the structural transition takes place; in fact symmetry breaking is marked by the splitting of the tetragonal 110 peak into orthorhombic 020 + 200 diffraction lines. In the high statistic data this quite weak peak (~ 20% of the highest intensity peak) is about 10 times more intense than that obtained by acquiring a conventional full diffraction pattern and hence these data are much more sensitive to splitting/broadening effects that can affect the 110 peak. The dependence of the full width at half maximum (FWHM) of the 110 diffraction peak on temperature was evaluated in two steps: first a standard pattern with no sample broadening was refined in order to determine the instrumental FWHM parameters and a resolution file was created; then the experimental FWHM parameter was calculated for 110 thermo-diffractograms. In this way the Lorentzian broadening of the reflection originated by the lattice microstrain was determined as a function of temperature.

Neutron powder diffraction analysis (NPD) of $(La_{1-y}Y_y)FeAsO$ samples ($y$ = 0.10, 0.20, 0.30) was carried out at the Institute Laue Langevin (Grenoble – France). Thermo-diffractograms were acquired on heating in continuous scanning mode in the $T$ range 2 – 200 K using the high intensity D1B diffractometer ($\lambda$ = 2.52 Å) to evaluate $T_{T-O}$, whereas high resolution NPD patterns for Rietveld refinement were collected at selected temperatures between 10 K and 300 K using the D1A diffractometer ($\lambda$ = 1.91 Å); structural and magnetic analyses as well as other details are reported elsewhere.[19] Even in this case refinements were carried out by applying a proper instrumental resolution file.

## 3. Results and discussion

The analysis of the high statistic 110 peak thermodiffractograms of the $SmFeAs(O_{1-x}F_x)$ series evidences that resolved splitting is observed at 90 K for the pure SmFeAsO sample, whereas in the F-substituted ones only peak broadening can be detected. In any case the onset of the structural transition is marked by an abrupt increase of the FWHM on cooling, that for undoped SmFeAsO takes place at $T_{T-O}$ ~ 175(10) K.

Figure 2 shows the evolution of the FWHM of the tetragonal 110 peak as a function of temperature for $SmFeAs(O_{0.95}F_{0.05})$ and $SmFeAs(O_{0.90}F_{0.10})$ samples. These samples are representative of the two different ground states exhibited by 1111-type compounds, the former accommodating a magnetic ordering below $T_M \sim 90$ K, the latter being superconductive ($T_c \sim 41$ K) with a completely suppressed static magnetic ordering, as revealed by μ-SR measurements[23] and resistive measurements (Figure 1); they are thus located in different regions of the electronic phase diagram (inset of Figure 2).[7] It is evident that above ~180 K the width of the peak homogeneously evolves, whereas around ~160 - 170 K an abrupt increase takes place in both samples; note that this is the only anomaly affecting the peak shape in the inspected temperature range, thus marking the tetragonal to orthorhombic transition. For the pure sample the same anomaly is observed at about the same temperature (~175 K). In a similar way $T_{T-O}$ was previously evaluated in Ba(Fe$_{1-x}$Cr$_x$)$_2$As$_2$,[25] whereas a significant broadening of the 110 peak at 180 K was detected for LaFeAsO with a better fit of the orthorhombic structural model below this temperature.[26]

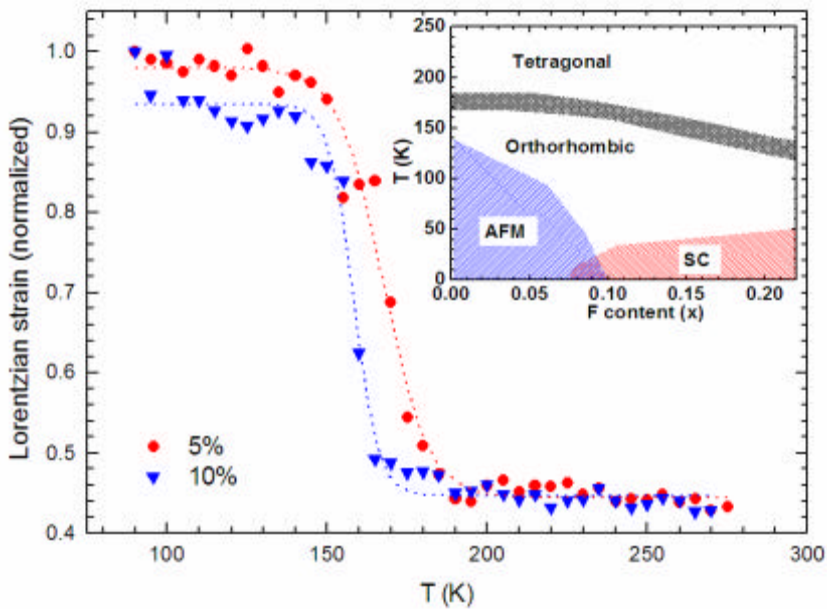

Figure 2: Evolution with temperature of the Lorentzian broadening for the tetragonal 110 peaks of SmFeAs(O$_{0.95}$F$_{0.05}$) and SmFeAs(O$_{0.90}$F$_{0.10}$) (normalized data; dotted lines are sigmoidal fits of the data); the inset shows the revised phase diagram of the SmFeAs(O$_{1-x}$F$_x$) system.[7]

The evolution of the FWHM of the 110 peak on cooling is qualitatively similar for all the samples, suggesting that F-substitution does not prevent the tetragonal to orthorhombic phase transition, but slightly decreases $T_{T-O}$.

Taking into account the results obtained by the FWHM analysis, Rietveld refinements were carried out assuming a tetragonal structural model ($P4/nmm$ structural model) for $T > 170$ K; in the temperature region where the 110 peak is broadened (~ below 170 K) the refinements were carried out using both a tetragonal and an orthorhombic structural model ($Cmme$ structural model). As a result an actual improvement of the fit was obtained using the orthorhombic model, whatever the F content. Structural data at 290 K and 90 K are reported in Table 1 and 2, respectively, whereas Figure 3 shows the Rietveld refinement plot of SmFeAs($O_{0.95}F_{0.05}$) selected as representative (data collected at 90 K).

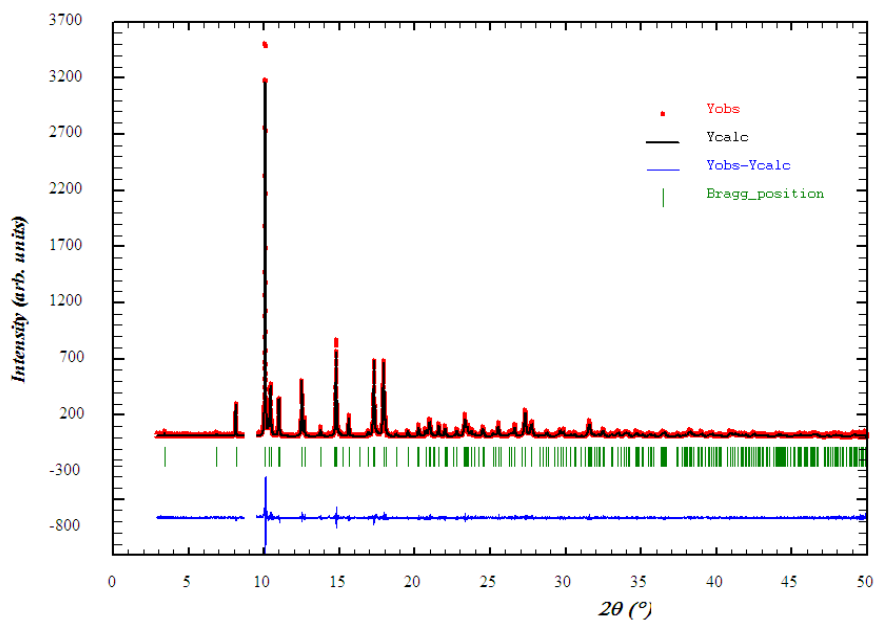

Figure 3: Rietveld refinement plot for SmFeAs($O_{0.95}F_{0.05}$) (data collected at 90 K). The observed intensity data are plotted in the upper field as points; the calculated pattern is the superposed solid-line curve; the short vertical bars in the middle field indicate the positions of the possible Bragg reflections; the difference is shown in the lower field; in the excluded region a faint peak of ice (condensed on the sample holder on cooling) was present.

In Figure 4 is reported the evolution of the cell parameters between 90 and 290 K (orthorhombic $a$ and $b$ values are divided by $\sqrt{2}$), where the progressive decrease of the orthorhombic distortion with the increase of F content can be appreciated (inset of the left panel). This effect can mask the symmetry breaking when the F content is relatively high; in this case the orthorhombic distortion can be hardly appreciated, revealed only by selective peak broadening whose amplitude strongly depends on instrumental and analytical conditions. On the basis of these results and μ-SR analysis a new phase diagram for the SmFeAs($O_{1-x}F_x$) system has been previously drawn[7] and is reported in the inset of Figure 2.

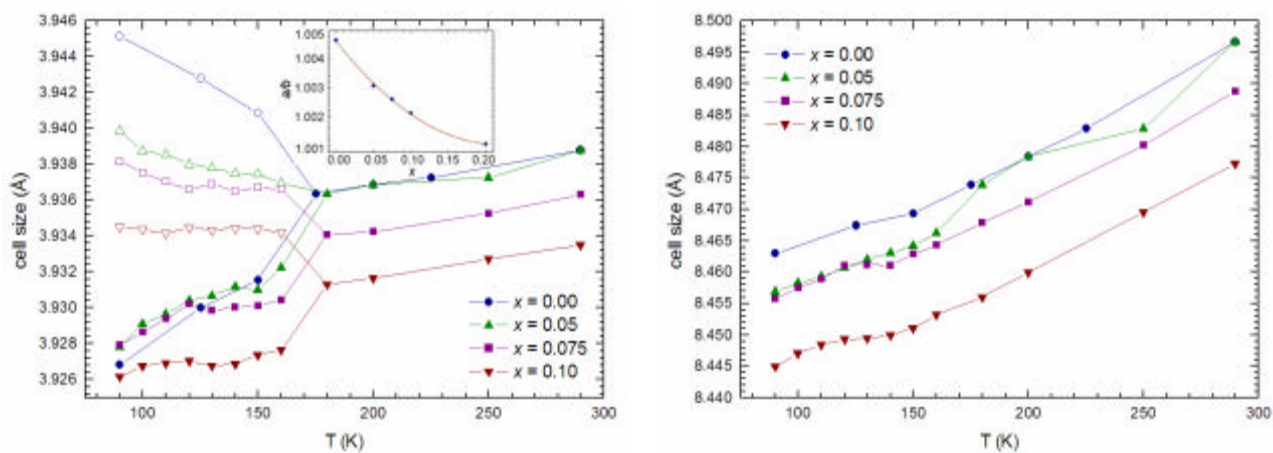

Figure 4: Dependence of the cell parameters on $T$ for the analyzed SmFeAs($O_{1-x}F_x$) compounds. Panel on the left: evolution of the $a$ (full symbol) and $b$ (empty symbol) cell parameters on $T$ (in the orthorhombic phase field both values were divided by $\sqrt{2}$); the inset shows the evolution of the orthorhombic distortion with F content. Panel on the right: evolution of the cell parameter $c$ on $T$. Error bars are smaller than the symbols; lines are guide to eyes.

The analysis of the refined anisotropic strain parameters in the tetragonal structural model of SmFeAsO reveals that microstrain is present mainly along $<hh0>$ and appreciably increases on cooling down to $T_{T-O}$ ~ 175 K. Below 175 K the refinements carried out applying the orthorhombic structural model reveal no remarkable microstrain contribution to the diffraction patterns.

In order to determine the possible driving mechanism responsible for the symmetry breaking, micro-structural evolution as a function of temperature was investigated by refining the anisotropic strain parameters and analyzing the broadening of diffraction lines by means of the Williamson-Hall plot method.[27] Generally, in the case where size effects are negligible and the microstrain is isotropic, a straight line passing through all the points and through the origin has to be observed, where the slope provides the microstrain: the higher the slope the higher the microstrain. When size contribution (size of the coherently diffracting domain) is not negligible, this straight line intersects the *y* axis at a value $1/t$, where $t$ is the size of the domain. So in these plots the slope provides the microstrain contribution whereas the reciprocal of the intercept an estimate of the size. If the broadening is not isotropic, size and strain effects along some crystallographic directions can be obtained by considering different orders of the same reflection.

In our samples integral breadths values ($\beta^* = A/I$, where *A* and *I* are the area and the height of the peak, respectively) for different order of a same reflection lie on a straight line passing through the origin, indicating that the size contribution is negligible, whereas the slope of the lines varies systematically with lattice direction, due to an anisotropic microstrain broadening. In particular the higher level of microstrain is exhibited by the *hh*0 lines and increases on cooling, raising up just above the structural transition. This behaviour is peculiar of the *hh*0 lines as well as of the diffracting peaks with a strong component along <*hh*0>. In fact the microstrain contribution maintains almost constant in the whole inspected *T* range for the *h*00 and 00*l* lines. In Figure 5 the Williamson-Hall plots obtained from SmFeAsO data collected at 290 K (empty symbols) and 175 K (full symbols) are superposed and illustrate such a behaviour (note that the *h*00 data have almost exactly the same values at both temperatures).

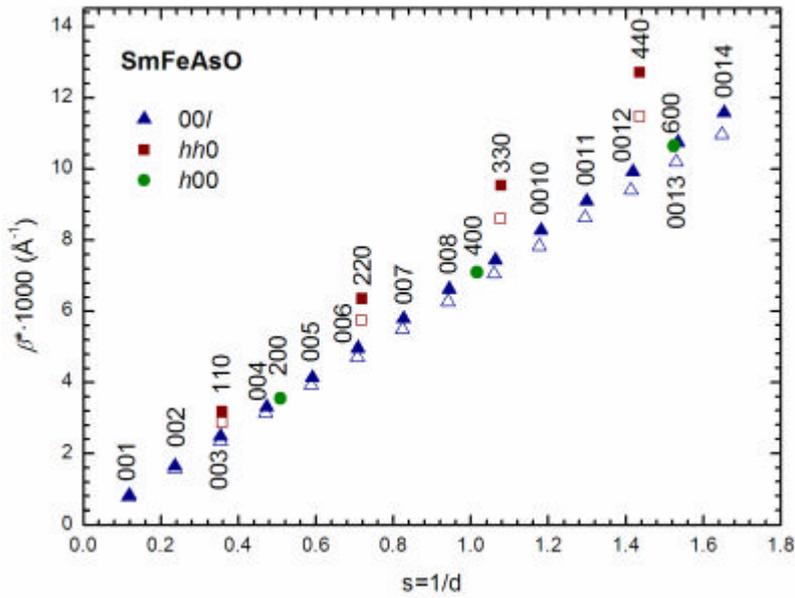

Figure 5: Superposition of the Williamson-Hall plots obtained for the SmFeAsO sample using the diffraction data collected at 290 K (empty symbols) and 175 K (full symbols); for clarity different orders of the 00$l$, $hh$0 and $h$00 reflections only are reported.

The detailed evolution of the microstrain along <$hh$0> can be followed by plotting the integral breadth of a $hh$0 diffraction line as a function of $T$. This is illustrated in the left panel of Figure 6 for the 110 line of SmFeAsO: as the temperature decreases microstrain increases reaching a maximum value just above $T_{T-O}$; at lower temperature, with the occurrence of the orthorhombic structure, microstrain is suppressed. Thus lattice microstrain is indicative of a phenomenon that leads to the progressive destabilization of the tetragonal structure in favour of the orthorhombic symmetry.

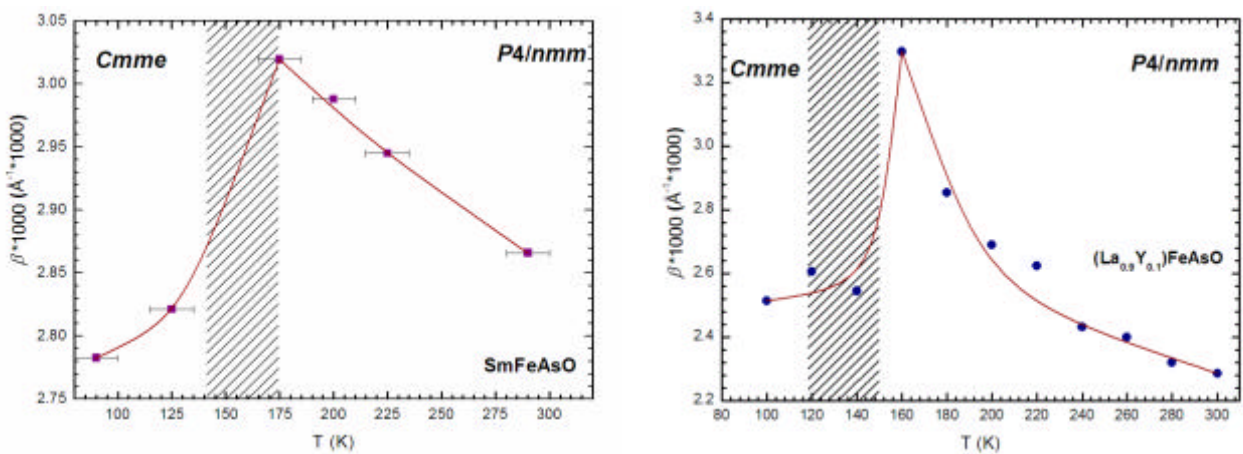

Figure 6: Evolution with temperature of the microstrain affecting the 110 peak in SmFeAsO (from SRPD data, on the left) and 220 peak in (La$_{0.9}$Y$_{0.1}$)FeAsO (from NPD data, on the right); lines are guide to eyes.

It is worth to note that the curves plotted in Figure 2 are obtained by a mere evaluation of the Lorentzian component of a single peak FWHM; conversely the curves plotted in Figure 6 are obtained by an actual microstructural analysis, where line breadths is measured by the more reliable integral breadth and the whole diffraction pattern is analyzed. As a consequence this analysis is much more sensitive and in fact microstrain broadening can be appreciated above structural transition.

Exactly the same evolution of the microstrain along $\langle hh0\rangle$ is found in the $(La_{1-y}Y_y)FeAsO$ system; Figure 7 shows the Rietveld refinement plot for $(La_{0.90}Y_{0.10})FeAsO$, selected as representative and obtained using NPD data collected at 10 K; the structural, magnetic and resistive properties of this system are detailed in some previous works.[19,21]

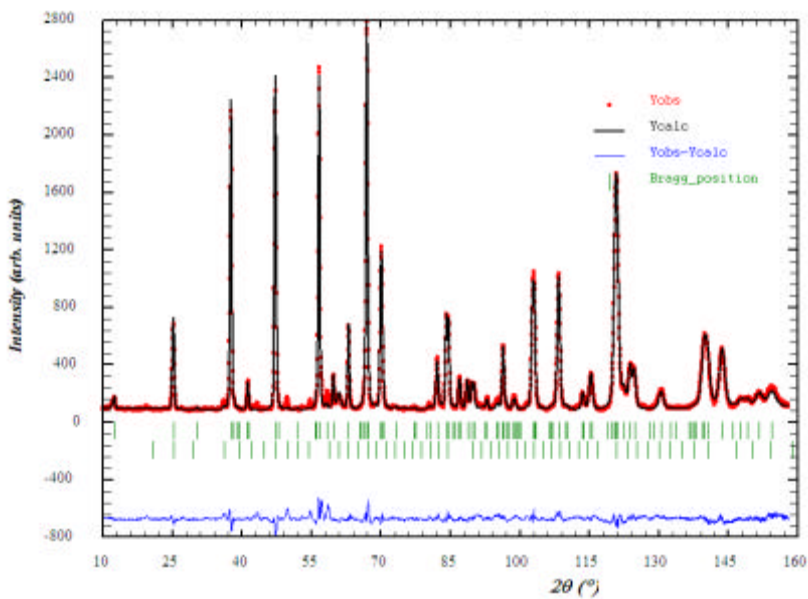

Figure 7: Rietveld refinement plot for $(La_{0.90}Y_{0.10})FeAsO$ (data collected at 10 K). The observed intensity data are plotted in the upper field as points; the calculated pattern is the superposed solid-line curve; the short vertical bars in the upper and lower middle field indicate the positions of the possible Bragg reflections for $(La_{0.90}Y_{0.10})FeAsO$ and secondary $Y_2O_3$ (< 5%); the difference is shown in the lower field.

In $(La_{0.90}Y_{0.10})FeAsO$ an increase of the microstrain along $\langle hh0\rangle$ takes place on cooling, that reaches its maximum value just above $T_{T-O}$; at lower temperature is abruptly suppressed with the occurrence of the orthorhombic structure (Figure 6, right panel). In this case the analysis of the 220 diffraction line is carried out using high resolution NPD data (where the 110 peak intensity is negligible) and hence microstrain is not electronic in origin (distortion of the electron shell), but reflects an effective displacement of the atomic nuclei. Generally speaking microstrain is originated by local fluctuations of the interplanar spacing; hence its occurrence in the tetragonal phase reveals a local breaking of this symmetry in favour of the orthorhombic one and this effect increases as $T_{T-O}$ is approached from above. Possibly these microstrains induces the anisotropic physical properties generally ascribed to nematic correlations occurring in the tetragonal phase.

When chemical substitution is carried out it is fundamental to obtain a homogenous distribution of the substituting element within the phase, in our case F within $SmFeAs(O_{1-x}F_x)$. Line broadening analysis by means of the Williamson-Hall plot method can also provide information on compositional homogeneity; in fact variations of lattice parameters can occur among crystallites characterized by different F content, originating line broadening due to a superposition of sub-line profiles with different positions. In such a case, within the tetragonal system, peak broadening depends on the square cosine of the angle $j$ between the diffraction vector and the $c$ direction;[28] hence an increase of the peak breadth should be observed with the decrease of the angle $j$ when compositional homogeneity is not obtained. Figure 8 reveals that such a kind of broadening is negligible in $SmFeAs(O_{0.90}F_{0.10})$, since the slope of the straight line fitting the 00$l$ reflections is lower than the corresponding line fitting the $h$00 reflections, and its slope is about the same measured in SmFeAsO. As a conclusion F is homogeneously distributed within the sample and then the abrupt increase of the FWHM signing the symmetry breaking cannot related to possible compositional variations within the sample, that is to regions of the sample where the F content is strongly decreased (de-mixed sample).

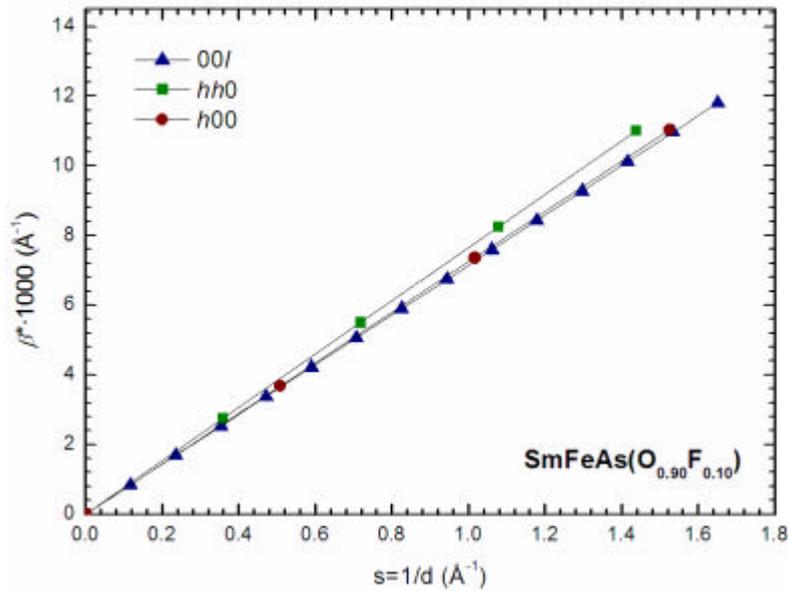

Figure 8: Williamson-Hall plot for SmFeAs($O_{0.90}F_{0.10}$) (data collected at 290 K).

As previously discussed a strong increase along <hh0> in the tetragonal structure is detected above $T_{T-O}$; Figure 9 shows the relationship between the tetragonal and the orthorhombic structures (both viewed along [001]) and it is evident that tetragonal <hh0> lattice microstrain is along the near-neighbour Fe atoms. Conversely no microstrain develops along <h00>, that is along next-near-neighbour Fe atoms. In the tetragonal phase each Fe atom is surrounded by four near-neighbour Fe atoms at the same distance; with the occurrence of the orthorhombic structure Fe-Fe bond lengths branch, as illustrated in Figure 9. Hence lattice microstrains detected in the tetragonal phase are the signature of a tendency of Fe-Fe bond lengths to branch.

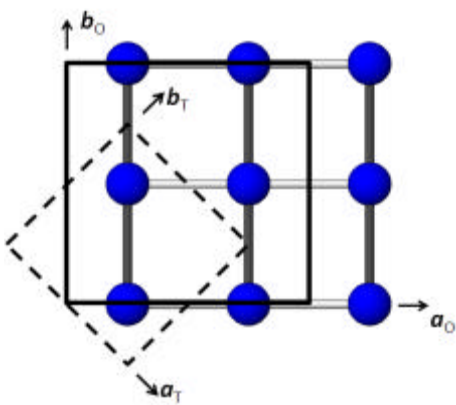

Figure 9: Relationship between the tetragonal and the orthorhombic structure of 1111-type compounds along [001]; spheres indicates the position of Fe atoms in the cell, whereas subscripts T and O stands for tetragonal and orthorhombic, respectively. Darker and lighter sticks represent longer and shorter Fe-Fe bonds in the orthorhombic structure, respectively.

As illustrated in a previous investigation,[19] in $(La_{1-y}Y_y)FeAsO$ compounds the evolution of $(a_o + b_o)/2$ ($a_o$ and $b_o$ are the orthorhombic cell parameters) below $T_{T\text{-}O}$ is continuous with that observed at higher temperature for the tetragonal cell parameters $a_T$. This feature is typical of order-disorder phase transitions and is observed, for example, in manganites undergoing an orbital ordering due to Jahn-Teller distortion.[29] On the other hand several features characterizing the symmetry breaking in 1111-type compounds comply with orbital ordering. First of all at the structural transition the analyzed $SmFeAs(O_{1-x}F_x)$ compounds show no volume discontinuity as a function of $T$. This is a common feature of 1111-compounds, as illustrated in Figure 10, showing the evolution of the volume around $T_{T\text{-}O}$ for both SmFeAsO and $(La_{0.9}Y_{0.1})FeAsO$, and is typical of structural transitions where orbital order takes place. A faint kink is displayed at $T_M$ by both systems, originated by magnetistriction. Such a behaviour deeply differs from what is observed in $Fe_{1+y}Te$,[30] where a net discontinuity accompanies symmetry breaking, suggesting that the structural transition is driven by different forces in 1111- and 11-type compounds.

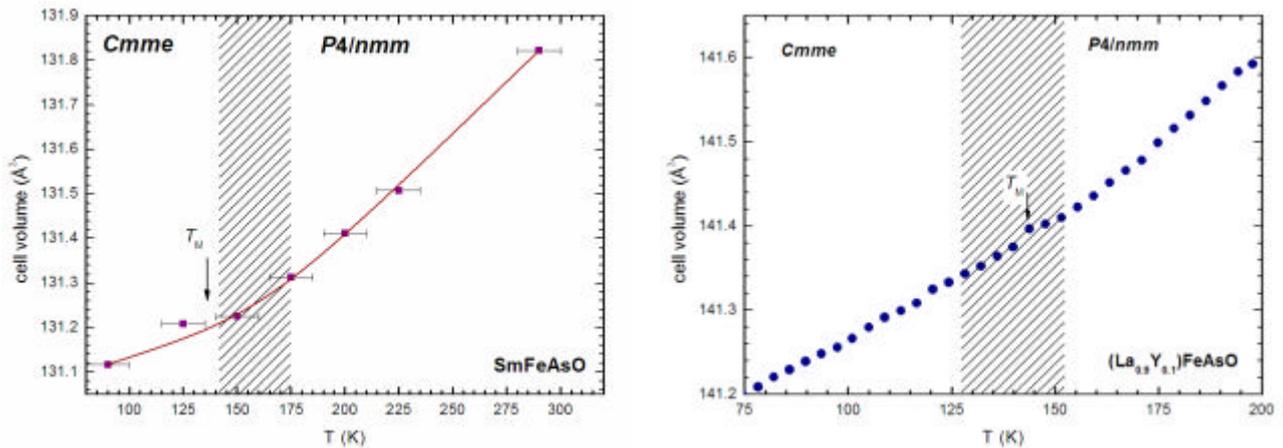

Figure 10: Evolution of the volume as a function of $T$ for SmFeAsO and $(La_{0.9}Y_{0.1})FeAsO$.

Symmetry breaking is a first order transition in 1111-type compounds, a feature in agreement with orbital ordering. The first order nature of the transition is evidenced in our diffraction data by the coexistence of both polymorphic phases within a specific temperature range. Clear evidence for this can be appreciated in Figure 11, showing an enlarged view of NPD patterns of $(La_{0.9}Y_{0.1})FeAsO$, selected as representative, collected at different temperatures; this sample exhibits an onset of symmetry breaking at 153(2) K, whereas at 126(2) K the transition is completed.[19] At 300 K only the tetragonal 322 line is present, whereas at 100 K it splits into the orthorhombic 152 and 512 lines. Data collected at 140 K, in the middle of the transition, clearly evidence the presence of both tetragonal and orthorhombic diffraction lines, implying phase coexistence; further, Rietveld refinement indicates that the molar percentage of the two polymorphs is about the same at this temperature. The rather large transition temperature width and phase coexistence are both indicative of a first order nature of the phenomenon, consistent with previous findings.[31,32]

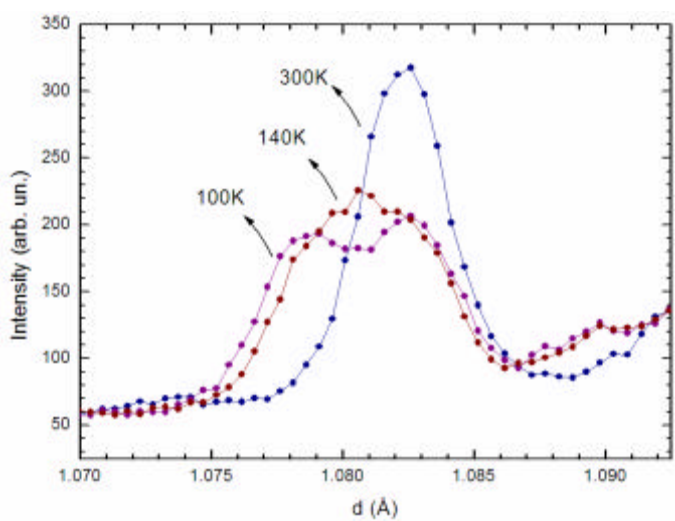

Figure 11: Splitting of the tetragonal 322 diffraction peak into the orthorhombic 152 and 512 lines with temperature in $(La_{0.9}Y_{0.1})FeAsO$.

Finally in the case of orbital ordering a group – subgroup relationship must hold between the disordered and the ordered structure and this is exactly the case for the symmetry breaking in 1111-type compounds, where the ordered orthorhombic structure (*Cmme* space group) is a subgroup of the disordered tetragonal structure (*P4/nmm* space group).

These results suggest a scenario where the lattice microstrain observed in the tetragonal phase is originated by the tendency of Fe orbitals to order along <*hh*0>; this phenomenon locally breaks the tetragonal symmetry above $T_{T-O}$ and produces nematic correlations between Fe orbitals, determining anisotropic physical properties. In this context the tendency towards orbital ordering in the orbital-disordered tetragonal phase properly reflects the 'order form disorder' effect claimed for nematic correlations; order is driven by orbital physics, as suggested by some recent theoretical investigations.[17,18] Then the stability of the tetragonal structure is progressively reduced on cooling, down to $T_{T-O}$, where orbital ordering percolates forming the orthorhombic structure and microstrain is suppressed.

Lattice microstrain and symmetry breaking takes place even when static magnetic order is suppressed, in agreement with theoretical investigations stating the primary role played by orbital ordering on the in-plane anisotropy and symmetry breaking as well.[17]

In conclusion several structural and microstructural features suggest the occurrence in the tetragonal phase of nematic fluctuations induced by the tendency towards ordering of Fe orbitals, whose amplitude progressively raises as $T_{T-O}$ is approached from above. Thus orbital ordering is found to play a primary role in the symmetry breaking mechanism of these compounds.

## 4. Conclusions

A careful microstructural analysis reveals that the first order tetragonal to orthorhombic structural transition is not suppressed in SmFeAs($O_{1-x}F_x$), even at optimal doping. An abrupt broadening of the full width at half maximum of the 110 diffraction peak marks symmetry breaking. Orthorhombic distortion is decreased but not suppressed by the increase of F content. Lattice microstrain along <hh0> are detected in the tetragonal phase in both SmFeAs($O_{1-x}F_x$) and ($La_{1-y}Y_y$)FeAsO compounds, whose amplitude increases as the structural transition temperature is approached from high tempertaure. This lattice microstrain is probably related to a progressive tendency of 3$d$ Fe orbitals to order (orbital order); after symmetry breaking, lattice microstrain is suppressed in the orthorhombic phase. Hence experimental evidences suggest that the structural transition in 1111-type compounds is likely driven by orbital ordering, in agreement with some recent theoretical calculations.

**Acknowledgments**

A.M. acknowledges H. Emerich of the SNBL at ESRF (Grenoble) and C. Ritter of the ILL (Grenoble) for their support during experiments. This work has been supported by FP7 European projects SUPER-IRON (n°283204).

Table 1: Structural data at 290 K for SmFeAs($O_{1-x}F_x$) samples obtained by Rietveld refinement (*P*4/*nmm* space group; Sm at 2*c*; Fe at 2*b*; As at 2*c*; O/F at 2*a*).

|   | SmFeAsO | SmFeAs($O_{0.95}F_{0.05}$) | SmFeAs($O_{0.925}F_{0.075}$) | SmFeAs($O_{0.90}F_{0.10}$) |
|---|---|---|---|---|
| *a* [Å] | 3.9389(1) | 3.9374(1) | 3.9365(1) | 3.9335(1) |
| *c* [Å] | 8.4969(1) | 8.4912(1) | 8.4887(1) | 8.4774(1) |
| *z* Sm | 0.1369(1) | 0.1384(1) | 0.1385(1) | 0.1400(1) |
| *z* As | 0.6599(2) | 0.6606(2) | 0.6604(2) | 0.6611(2) |
| $R_F$ | 4.77 | 4.47 | 4.35 | 4.29 |
| $R_B$ | 4.66 | 3.15 | 3.98 | 4.08 |

Table 2: Structural data at 90 K for SmFeAs(O$_{1-x}$F$_x$) samples obtained by Rietveld refinement (*Cmme* space group; Sm at 4$g$; Fe at 4$b$; As at 4$g$; O/F at 4$a$).

|  | SmFeAsO | SmFeAs(O$_{0.95}$F$_{0.05}$) | SmFeAs(O$_{0.925}$F$_{0.075}$) | SmFeAs(O$_{0.90}$F$_{0.10}$) |
|---|---|---|---|---|
| $a$ [Å] | 5.5539(1) | 5.5619(1) | 5.5611(1) | 5.5573(1) |
| $b$ [Å] | 5.5797(1) | 5.5658(1) | 5.5643(1) | 5.5603(8) |
| $c$ [Å] | 8.4638(1) | 8.4573(1) | 8.4558(1) | 8.4452(1) |
| $z$ Sm | 0.1372(1) | 0.1385(1) | 0.1388(1) | 0.1401(1) |
| $z$ As | 0.6596(2) | 0.6607(2) | 0.6604(1) | 0.6607(1) |
| $R_F$ | 4.39 | 4.42 | 4.37 | 3.26 |
| $R_B$ | 3.73 | 3.29 | 4.07 | 2.53 |

**References**

[1] Kamihara, Y.; Watanabe, T.; Hirano, M.; Hosono, H., J. Am. Chem. Soc., 130 (2008) 3296

[2] M. Fratini, R. Caivano, A. Puri, A. Ricci, Zhi-An Ren, Xiao-Li Dong, Jie Yang, Wei Lu, Zhong-Xian Zhao, L. Barba, G. Arrighetti, M. Polentarutti, A. Bianconi, Supercond. Sci. Technol. 21 (2008) 092002

[3] T Nomura, S W Kim, Y Kamihara, M Hirano, P V Sushko, K Kato, M Takata, A L Shluger, H Hosono, Supercond. Sci. Technol. 21 (2008) 125028

[4] S. Margadonna, Y. Takabayashi, M. T. McDonald, M. Brunelli, G. Wu, R. H. Liu, X. H. Chen, K. Prassides, Phys. Rev. B 79, 014503 (2009)

[5] A. Martinelli, A. Palenzona, C. Ferdeghini, M. Putti, H. Emerich, J. All. Comp. **477** (2009) L21

[6] Jun Zhao, Q. Huang, C. de la Cruz, Shiliang Li, J. W. Lynn, Y. Chen, M. A. Green, G. F. Chen, G. Li, Z. Li, J. L. Luo, N. L. Wang, Pengcheng Dai, Nature Materials 7, 953 (2008)

[7] A. Martinelli, A. Palenzona, M. Tropeano, M. Putti, C. Ferdeghini, G. Profeta, E. Emerich, Phys. Rev. Lett. 106, 227001 (2011)

[8] K. Kasperkiewicz, J.-W. G. Bos, A. N. Fitch, K. Prassides, S. Margadonna, Chem. Commun. 707 (2009) 707

[9] T. Yildirim, Phys. Rev. Lett. 101, 057010 (2008)

[10] F. Ma, Z.-Y. Lu, T. Xiang, Phys. Rev. B 78, 224517 (2008)

[11] S. A. J. Kimber, D. N. Argyriou, I. I. Mazin, arXiv:1005.1761

[12] C. Lester, J.-H. Chu, J. G. Analytis, T. G. Perring, I. R. Fisher, S. M. Hayden, Phys. Rev. B 81, 064505 (2010)

[13] S. Ishibashi, K. Terakura, H. Hosono, J. Phis. Soc. Jap. **77**, 053709 (2008)

[14] Chi-Cheng Lee, Wei-Guo Yin, Wei Ku, Phys. Rev. Lett. 103, 267001 (2009)

[15] F. Krüger *et al.*, Phys. Rev. B 79, 054504 (2009)

[16] Weicheng Lv, Jiansheng Wu, P. Phillips Phys. Rev. B 80, 224506 (2009)

[17] Weicheng Lv, P. Phillips Phys. Rev. B 84, 174512 (2011)

[18] M. S. Laad, L. Craco, Phys. Rev. B 84, 054530 (2011)

[19] A. Martinelli, A. Palenzona, M. Tropeano, C. Ferdeghini, M. R. Cimberle, C. Ritter, Phys. Rev. B 80, 214106 (2009)

[20] A. Martinelli, M. Ferretti, P. Manfrinetti, A Palenzona, M. Tropeano, M R. Cimberle, C. Ferdeghini, R. Valle, C. Bernini, M. Putti, A. S. Siri. Supercond. Sci. Technol. 21 (2008) 095017

[21] M. Tropeano, C. Fanciulli, F. Canepa, M. R. Cimberle, C. Ferdeghini, G. Lamura, A. Martinelli, M. Putti, M. Vignolo, A. Palenzona, Phys. Rev. B 79, 174523 (2009)

[22] M. Tropeano, I. Pallecchi, M.R. Cimberle, C. Ferdeghini, G. Lamura, M. Vignolo, A. Martinelli, A. Palenzona, M. Putti, Superconductor Science and Technology 23 (2010) 054001

[23] S. Sanna, R. De Renzi, G. Lamura, C. Ferdeghini, A. Palenzona, M. Putti, M. Tropeano, T. Shiroka, Phys. Rev. B 80 (2009) 052503

[24] J. Rodríguez–Carvajal Physica B 192, 55 (1993)

[25] K. Marty, A. D. Christainson, C. H. Wang, M. Matsuda, H. Cao. L. H. VanBebber, J. L. Zarestky, D. J. Singh, A. S. Sefat, M. D. Lumsden, Physical Review B 83, 060509(R) (2011)

[26] M. A. McGuire, A. D. Christianson, A. S. Sefat, B. C. Sales, M. D. Lumsden, R. Jin, E. A. Payzant, D. Mandrus, Y. Luan, V. Keppens, V. Varadarajan, J. W. Brill, R. P. Hermann, M. T. Sougrati, F. Grandjean, G. J. Long, Physical Review B 78 094517

[27] J. I. Langford, D. Louër, E. J. Sonneveld, J. W. Visser, Powder Diffr. 1 (1986) 211

[28] A. Leineweber, E. J. Mittemeijer, J. Appl. Cryst. (2004). 37, 123

[29] P. G. Radaelli, D. E. Cox, M. Marezio, and S.-W. Cheong, Phys. Rev. B 55, 3015 (1997)

[30] A. Martinelli, A. Palenzona, M. Tropeano, C. Ferdeghini, M. Putti, M. R. Cimberle, T. D. Nguyen, M. Affronte, C. Ritter, Physical Review B 81 (2010) 094115

[31] S. A. J. Kimber, D. N. Argyriou, F. Yokaichiya, K. Habicht, S. Gerischer, T. Hansen, T. Chatterji, R. Klingeler, C. Hess, G. Behr, A. Kondrat, B. Büchner, Phys. Rev. B 78, 140503(R) (2008)

32 J. W. Lynn, Pengcheng Dai, Physica C 469 (2009) 469